\documentstyle[12pt]{article}
\begin{document}
{\pagestyle{empty}

{\Large\begin{center}
{\bf MASS GENERATION IN THE STANDARD MODEL \\
WITHOUT DYNAMICAL HIGGS FIELD}
\end{center}}
\vskip 1cm

\begin{center}
{\bf Marek Paw\l owski}
\\
Soltan Institute for Nuclear Studies, Warsaw, POLAND
\\
{\bf and }
\\
{\bf Ryszard R\c aczka}
\\
Soltan Institute for Nuclear Studies, Warsaw, POLAND
\\
and
\\
Interdisciplinary Laboratory for Natural and Humanistic Sciences
\\
International School for Advanced Studies (SISSA), Trieste, ITALY
\\ ~ \\
{17 March 1994}
\end{center}

\vskip -12cm
\rightline{ILAS 2/94}
\vskip 12cm

\bigskip\bigskip
\vskip .5cm
\centerline{\bf Abstract}

\bigskip

We call attention to the fact that the gauge symmetry $SU(3)\times
SU(2)_{_L}\times U(1)$ of the Standard Model can be easily and
naturally extended by the local conformal symmetry connected with the
possibility of choosing the local length scale.  Restricting the
admissible interactions to the lowest order conformal invariant
interactions one gets the unique form of total lagrangian.  It
contains all Standard Model fields and gravitational interaction.
Using the unitary and the conformal gauge conditions we can eliminate
all four real components of the Higgs doublet in this model.  However
the masses of vector mesons, leptons and quarks are automatically
generated and are given by the same formulas as in the conventional
Standard Model.  In this manner one gets the mass generation without
the mechanism of spontaneous symmetry breaking and without the
remaining real dynamical Higgs field.

\vfil

\eject
{}~

\eject}

The Higgs mechanism of spontaneous symmetry breaking (SSB) provides a
simple and effective instrument in the Standard Model (SM) for mass
generation of weak gauge bosons, quarks and leptons. However, despite
of many efforts of several groups
of experimentalists \cite{LEP} the postulated Higgs
particle of the SM was not observed. Hence one might expect that there
may exist an alternative mechanism of mass generation in which the dynamical
Higgs field is absent.

We present in this work a new frame--work for gauge symmetry breaking and
mass generation. In the proposed formalism the conventional
gauge symmetry is extended by the local conformal symmetry. The breaking
of the extended symmetry by the conformal and the unitary gauge conditions
generates the masses
of weak
gauge bosons, leptons and quarks.

\medskip
Let $M^{3,1}$ be the pseudo--Riemannian space time with the metric
$g_{_{\alpha
\beta}}$ with the signature (+,-,-,-). Let $\Omega(x)$ be a smooth strictly
positive function on $M^{3,1}$. Then the conformal transformation in $M^{3,1}$
is defined as the transformation which changes the metric by the formula
$$g_{_{\alpha\beta}}(x)\rightarrow \tilde{g}_{_{\alpha\beta}}(x)=\Omega^2(x)
g_{_{\alpha\beta}}(x). \eqno (1)$$
The set of all conformal transformations forms the multiplicative abelian
infinite--dimensional group $C$ with the obvious group multiplication law.

Let $\Psi$ be a tensor or spinor field of arbitrary spin. A field equation of
$\Psi$ is called conformal invariant if there exist a number $s\in R$ such
that $\Psi(x)$ is a solution with the metric $g_{_{\alpha\beta}}(x)$ if and
only if
$$\tilde{\Psi}(x)=\Omega^s(x)\Psi(x) $$
is a solution with the metric $\tilde{g}_{_{\alpha\beta}}(x)$. The number $s$
is
called the conformal weight of $\Psi$ \cite{birel}, \cite{wald}, \cite{casta}.

It is known that the Maxwell and the Yang--Mills field strength
$F_{_{\alpha\beta}}$ and $F^{^a}_{_{\alpha\beta}}$ respectively has
the conformal degree $s_{_F}=0$, the massless Dirac field has the
conformal degree $s_{_\psi}=-{3\over 2}$ and the scalar
massless
field $\Phi$ satisfying Laplace--Beltrami equation
$$\triangle\Phi=0 $$
is not conformal invariant. In fact it was discovered by
Penrose that one has to add to the Lagrangian on $(M^{3,1},g)$ the term
$$ -{1\over 6}R\Phi^2 $$
where $R$ is the Ricci scalar, in order that the corresponding
field equation is conformal invariant with the conformal weight $s_{_\Phi}=-1$
\cite{penrose}.\medskip

We postulate that the theory of strong, electro--weak and gravitational
interactions is invariant under the group $G$
$$ G=SU(3)\times SU(2)_{_L}\times U(1)\times C \eqno (2)$$
where $C$ is the local conformal group defined by (1).
Let $\Psi$ be the collection of vector meson, fermion and scalar fields which
appear in the conventional minimal SM for electro--weak and strong
interactions.
Then the conformal and $SU(3)\times SU(2)_{_L}\times U(1)$ --gauge invariant
total lagrangian $L(\Psi)$ containing the lowest order conformal invariant
interactions is unique and has the form:
$$ L = [L_{_{G}}+L_{_{F}}+L_{_Y}+
L_{_\Phi}
- {1\over 6}\Phi^{\dagger}\Phi R  +
L_{_{grav}} ] \sqrt{-g} \eqno (3)
$$
Here $L_{_{G}}$ is the total lagrangian for the gauge fields $A^{^a}_{_\mu}$,
$W^{^b}_{_\mu}$ and $B_{_\mu}$, $a=1,...,8$, $b=1,2,3$ associated with
$SU(3)\times SU(2)_{_L}\times U(1)$ gauge group; $L_{_{F}}$ is the
lagrangian for fermion field interacting with the gauge fields; $L_{_\Phi}$
is the lagrangian for the scalar fields
$$L_{_\Phi}=(D\Phi)^{\dagger}(D\Phi) - \lambda(\Phi^{\dagger}\Phi)^2 \eqno
(4)$$
where $D$ denotes the covariant derivative with connections of all symmetry
groups;  $L_{_Y}$  represents  the  Yukawa  interactions  of fermion and
scalar
fields; the term $- {1\over 6}\Phi^{\dagger}\Phi R $ is the Penrose term
which assures that the lagrangian (3) is conformal invariant and
$$ L_{_{grav}} = -\rho C^2, \hskip1cm\rho>0, \eqno (5)$$
where $C_{\alpha\beta\gamma}^\delta$ is the Weyl tensor which is  conformally
invariant. Using the Gauss--Bonnet identity we can write $C^2$ in the form
$$C^2=2(R^{^{\mu\nu}}R_{_{\mu\nu}}-{1\over 3}R^2). \eqno (6)$$

It was shown by  Stelle \cite{Stelle} that quantum gravity defined by
(5) is perturbatively renormalizable whereas the quantum gravity defined by
the
Einstein lagrangian
$$L=\kappa^{-2} R\sqrt{-g}, \hskip1cm \kappa^2=(16\pi G)  \eqno (7)$$
coupled with matter is nonrenormalizable \cite{deser}.
Notice also that conformal symmetry implies that all coupling constants in the
present model are dimensionless.

The theory given by (3) is our conformally invariant proposition alternative
to the standard
Higgs--like theory with SSB.
Its new, most important feature is the
local conformal
invariance. It means that simultaneous rescaling of all fields (including
the field of metric tensor) with a common, arbitrary, space--time dependent
factor $\Omega(x)$ taken with a proper power for each field
(the conformal weight)
will leave the Lagrangian (3) unaffected. The symmetry has a clear and
obvious physical meaning \cite{narlikar}, \cite{wald}. It changes in every
point of the space--time
all dimensional quantities (lengths, masses, energy levels, etc) leaving
theirs ratios unchanged. It reflexes the deep truth of the nature that
nothing except the numbers has an independent physical meaning. It means
for example that we can prescribe arbitrary numbers to electron and proton
mass and that this prescription can differ from place to place in the space
and in the time. The only thing we have to care is to keep the ratio of
prescribed numbers consistent with experiment.

The freedom of choice of the length scale is nothing but the gauge fixing
freedom connected with the conformal symmetry group. In the
conventional approach we define the length scale in such a way that elementary
particle masses are the same for all times and in all places. This will be
the case when we rescale all fields with the x--dependent conformal factor
$\Omega(x)$ in such a manner that the length of scalar field doublet is fixed
i.e.

$$\Phi^{\dagger}\Phi={v^2 \over 2}=const. \eqno (8)
$$

Such a gauge fixing is distinguished by nothing but our convenience.
Obviously we can choose other gauge fixing condition, e.g. we can use the
freedom of conformal factor to set
$\sqrt{-g}=1$; this will lead to other local scales but will leave physical
predictions unchanged.

Inserting the gauge condition (8) into the Lagrangian (3) we obtain
$$ L^{gauged} = [L_{_{G}}+L_{_F}+L_{_\Phi}^{gauged}+L_{_Y}^{gauged}
- {1\over 12}v^2 R
 +  L_{_{grav}}
 ] \sqrt{-g},
 \eqno (9)
 $$
 in which the condition (8) was inserted into $L _{_\Phi}$ and $L_{_Y}$.
 This, together with an unitary gauge fixing of $SU(2)_{_L}\times U(1)$ gauge
 group, reduces the Higgs doublet to the form
 $$\Phi^{^{gauge}}={1\over\sqrt{2}}{0\choose v}, \hskip 1cm v>0 \eqno (10)$$
 and produce
 mass terms for some fermions and some vector bosons.
 Then the $\Phi$--lepton Yukawa interaction $L^l_{_Y}$
 $$L^l_{_Y}=-\sum_{i=e,\mu,\tau}G_{_i}\bar l_{_i}_R(\Phi^\ast l_{_i}_L)
 +h.c.$$
 passes into
 $$L^l_{_Y}^{^{gauged}}=-{1\over\sqrt{2}}v\sum_{i=e,\mu,\tau}G_{_i}\bar
 l_{_i}l_{_i} \eqno (11)$$
 giving the conventional, space--time independent lepton masses

 $$m_{_e}={1\over\sqrt{2}}G_{_e}v, \hskip1cm
 m_{_\mu}={1\over\sqrt{2}}G_{_\mu}v, \hskip1cm
 m_{_\tau}={1\over\sqrt{2}}G_{_\tau}v. \eqno (12)$$
 Similarly one generates from $\Phi$--quark Yukawa interaction $L^q_{_Y}$
 the corresponding quark masses. In turn from $L_{_\Phi}$-lagrangian (4)
 using the gauge condition (10) one obtains the $SU(2)$ - vector meson
 masses
 $$m_{_W}={v\over 2}g_2, \hskip1cm
 m_{_Z}={v\over 2}\sqrt{g_1^2+g^2_2} \eqno (13)$$
 where $g_1$ and $g_2$ are $U(1)$ and $SU(2)$ gauge coupling constants
 respectively. Then, analogously as in the case of conventional formulation
 of SM one can deduce the tree level relation between $v$ and $G_F$ --
 the four--fermion coupling constant of $\beta$--decay:
 $$v^2=(2G_F)^{-1}\rightarrow v=246GeV. \eqno (14)$$
 Here we have used the standard decomposition $g^{\mu\nu}\sqrt{-g}=
 \eta^{\mu\nu}+\kappa^\prime h^{\mu\nu}$ (see e.g. \cite{capper}) which reduces
 the tree level
 problem for the matter fields to the ordinary flat case task.

 We see therefore that the resulting expressions for masses of physical
 particles are identical as in the conventional SM.

 Let us stress that the conformal gauge condition like (8) does not break
 $SU(2)_{_L}\times U(1)$ gauge symmetry. The symmetry is broken (or rather
 one of gauge equivalent description is fixed) when (8) is combined with
 unitary gauge condition of electro--weak group leading to (10). However,
 also after imposing of a gauge condition like (10) we have a remnant of
 both the conformal and
 $SU(3)\times SU(2)_{_L}\times U(1)$ initial gauge symmetries: this is
 reflected in the special, unique  relations between couplings and
 masses of the gauged theory.
 \bigskip

 Notice that the final Lagrangian (9) contains the linear
 coupling with gravity $- {1\over 6}v^2 R$ which has the wrong, negative
 sign. This
 is the price for the positive sign of the kinetic energy of scalars in (4).
 We show now that
 this difficulty can be overcomed by a
 contribution from $L_{_{grav}}$. The role of $L_{_{grav}}$ is the following:
 the universe filled with matter is not necessarily flat. In order to describe
 it near some curved background we can expand the lagrangian (9) near some
 field configuration describing the background. It was done in the Appendix
 leading to the expanded lagrangian

 $$ L^{^{gauged}}= [L^{^{gauged}}_{_{matter}} +({1\over3}\rho R_{_0}-{1\over
 12}v^2) R
 -{1\over6}\rho R_{_0}^2]\sqrt{-g} + O(\epsilon^2)\eqno (15)$$
 where the empty, Einsteinian background is described by the metric $g_{_0}$
 and $\epsilon$ is the expansion parameter defined by (A1).
 The last term in parentheses is a constant which contributes to the
 cosmological constant.

 Setting now
 $${1\over3}\rho R_{_0}-{1\over 12}v^2=\kappa^{-2} = (16\pi G)^{-1}
 \eqno(16)$$
 we reproduce up to $O(\epsilon^2)$ terms the Einstein Lagrangian.
 We should mention that there were proposed also other methods of getting the
 Einsteinian term from the higher order gravitational interactions
 \cite{adler}, \cite{cheng}.
 \bigskip

 The presence of higher gravitational terms is also useful from the point of
 view
 of the renormalizability of the theory. The term (6) is nothing but the
 counterterm which appears when one tries to renormalize the ordinary Einstein
 gravity coupled with massless fields of spin $1/2$ and one
 \cite{deser}, \cite{duff}. It was shown in \cite{Stelle} that the theories
 with such terms are perturbatively renormalizable.

 Since the dynamical Higgs field can be gauged away in the proposed model
 one can afraid that
 the renormalizability of the matter sector is lost, especially the
 renormalizability
 of electro-weak part in the SM. However we must stress again that,
 in contrast to the
 ordinary case with massive nonabelian vector bosons, we start in our
 approach from the
 theory (3) whose the symmetry group is the direct product of the
 gauge symmetry group
 $SU(3)\times SU(2)_L\times U(1)$ and the conformal symmetry. Hence the action
 of the conformal group is independent on the action of the gauge
 groups. Consequently conformal gauge does not break the electro--weak
 symmetry as the condition (8) didn't. The mass formulas (12) and (13)
 were obtained when (8) was combined with the unitary gauge condition gauging
 $SU(2)_L\times U(1)$ group.

 In order to see the problem of renormalizability of the matter sector more
 clearly we may replace the gauge condition (8) with a
 different one which will be more suitable for the present analysis. When we
 choose the already mentioned unimodular condition
 $$\sqrt{-g}=1 \eqno (17)
 $$
 and we froze the rest of gravitational degrees of freedom demanding
 $$g^{\mu\nu}=\eta^{\mu\nu} \eqno (18)$$
 we will obtain from (3) the standard Higgs--like theory with the
 Higgs potential reduced to the quartic term only. This
 model is renormalizable by power counting theorem \cite{zavialov}.
 Hence we might expect that the model
 defined by (3) will be renormalizable.
 \bigskip\bigskip

 Thus we have obtained the desired phenomenon. Masses are generated without
 any SSB mechanism but due to the conformal and the unitary gauge fixing
 conditions. The
 vector
 meson masses were obtained
 from $L_{_\Phi}$ whereas the fermion masses were obtained from the Yukawa
 interaction
 $L_{_Y}$.
 The only remnant of the Higgs potential is the quartic
 ${\lambda\over 4}v^4$ term which
 contributes to
 the cosmological constant. We see therefore that
 for  the   minimal   SM   our   mechanism
 generates the same spectrum of masses of physical particles as
 in the case of SSB in Higgs theory. It also reproduce all
 fermion and gauge boson couplings except the coupling with scalar fields
 which were gauged away.

 It should be mentioned that there exist several
 proposals in the literature for a modification of SM by inclusion of
 gravitational interactions \cite{inni}. However these models are
 perturbatively nonrenormalizable in the
 gravitational sector or/and do not secure the proper Einsteinian limit.
 The main difference between these models and ours consists on
 the local conformal invariance
 of our model and the presence of the specific higher gravitational term
 $C^2$.

 Finally we note that the alternative
 mechanism of mass generation
 in SM implied by the causality condition of gravitational interactions
 was presented in \cite{fr}: in this model it remained however the dynamical
 real Higgs field.

 \bigskip

 \leftline{ACKNOWLEDGMENTS}
 \medskip

 The authors are grateful to Prof. P. Budinich and Prof. S. Fantoni for
 their kind
 hospitality at SISSA.

 \bigskip
 \leftline{APPENDIX}
 \medskip

 Let $g_{_0}$ be the background metric. Then, for small perturbation from
 $g_{_0}$ the full dynamical metric $g$ can be written down as

 $$g^{\mu\nu}=g_{_0}^{\mu\nu} + \epsilon h^{\mu\nu} \eqno(A1)$$
 where $\epsilon$ is a small parameter which will help us to control the
 expansion. Similarly we can expand the inverse metric
 $$ g_{\mu\nu}=g_{_0}_{\mu\nu} - \epsilon h_{\mu\nu}  + O(\epsilon^2)
 \eqno(A2)$$
 where $g_{_0}_{\mu\nu}$ is the inverse of $g_{_0}^{\mu\nu}$ and
 $$h_{\mu\nu} = h^{\alpha\beta}g_{_0}_{\mu\alpha}g_{_0}_{\nu\beta}. \eqno
 (A3)$$

 The curvature tensor can also be expanded in powers of $\epsilon$. For the
 Ricci tensor we get
 $$R^{\mu\nu}=R_{_0}^{\mu\nu} + \epsilon R_{_2}^{\mu\nu} + O(\epsilon^2)
 \eqno(A4)$$
 where $R_{_2}^{\mu\nu}$ is some contribution depending on $h$ and $g_{_0}$.

 For the scalar curvature we get
 $$R=R_{_0} +\epsilon(R_{_2}^{\mu}_{\mu} -R_{_0}^{\mu\nu}h_{\mu\nu})
 + O(\epsilon^2). \eqno(A5)$$
 and for the term $R^{\mu\nu}R_{\mu\nu}$ contributing to the square of Weyl
 tensor (6) we get
 $$R^{\mu\nu}R_{\mu\nu}=R_{_0}^{\mu\nu}R_{_0}_{\mu\nu} + 2\epsilon
 (R_{_2}^{\mu\nu}R_{_0}_{\mu\nu} -
 R_{_0}^{\mu\nu}h_{\mu\alpha}R_{_0}^{\alpha}_{\nu})+O(\epsilon^2), \eqno(A6)$$
 where the symmetry of the metric tensor was used.

 Now, it is natural to assume that vacuum
 Einstein equations holds for the background geometry.
 This implies that the Ricci tensor $R_{_0}^{\mu\nu}$
 is proportional to the metric $g_{_0}^{\mu\nu}$ and then
 $R_{_0}^{\mu\nu}=R_{_0}g_{_0}^{\mu\nu}/4$. With this assumption we get from
 (A6)
 $$R^{\mu\nu}R_{\mu\nu}={1\over 4}R_{_0}^2 + {1\over 2}R_{_0}\epsilon
 (R_{_2}^{\mu}_{\mu} - {1\over 4}R_{_0}h^{\mu}_{\mu}) +O(\epsilon^2)
 \eqno(A7)$$
 and from (A5) we obtain the relation
 $$\epsilon(R_{_2}^{\mu}_{\mu} - {1\over 4}R_{_0}h^{\mu}_{\mu}) =
 R-R_{_0} + O(\epsilon^2). \eqno(A8)$$
 Inserting (A8) into (A7) we get
 $$R^{\mu\nu}R_{\mu\nu}={1\over 4}R_{_0}^2 + {1\over 2}R_{_0}(R-R_{_0})
 +O(\epsilon^2). \eqno(A9)$$
 where the $\epsilon$--expansion of $R-R_{_0}$ begins with the linear term in
 $\epsilon$.
 Finally, because
 $$R^2=R_{_0}^2 +2R_{_0}(R-R_{_0}) + O(\epsilon^2), \eqno(A10)$$
 we obtain the following expression for the square of Weyl tensor

 $$C^2={1\over 6}R^2_{_0}-{1\over 3}R_{_0}R + O(\epsilon^2). \eqno(A11)$$
 Inserting this into the lagrangian (9) we get the formula (15).
 
 \clearpage


\bigskip

 \begin{thebibliography}{99}
 \bibitem{LEP} ALEPH Coll., Phys. Rep. {\bf 216} (1992) 253; DELPHI Coll., P.
Abreu
 {\sl et al.}, Nucl. Phys. {\bf B373} (1992) 3; L3 Coll., O. Adriani {\sl et
al.},
 Phys. Lett. {\bf 253B} (1993) 391; OPAL Coll., M. Akrawy {\sl et al.}, Phys.
Lett.
 {\sl 253B} (1991) 511.
 \bibitem{birel} N.D. Birrel, P.C.W. Davies, {\sl Quantum Fields in Curved
Space}
 (Cambridge Univ., Cambridge 1982).
 \bibitem{wald} R.M. Wald, {\sl General Relativity} (The Univ. of Chicago
Press,
 Chicago 1984).
 \bibitem{casta} M.A. Castagnino, J.B. Sztrajman, J. Mat. Phys. {\bf 27} (1986)
 1037.
 \bibitem{penrose} R. Penrose, Ann. of Phys. {\bf 10} (1960) 171.
 \bibitem{Stelle} K.S. Stelle, Phys. Rev. {\bf D16} (1977) 953
 \bibitem{adler} S.L. Adler, Rev. Mod. Phys. {\bf 54} (1982) 729.
 \bibitem{cheng} H. Cheng,Phys. Rev. Lett. {\bf 61} (1988) 2182.
 \bibitem{deser} S. Deser, P. von Nieuwenhuizen, Phys. Rev. {\bf D10} (1974)
 401.
 \bibitem{narlikar} J.V. Narlikar, {\sl Introduction to Cosmology} (Jones and
 Bartlet Pub. 1983).
 \bibitem{capper} D.M. Capper, G. Leibbrandt, M.R. Medrano, Phys. Rev. {\bf D8}
 (1973) 4320.
 \bibitem{duff} D.M. Capper, M.J. Duff, and L. Halpern, Trieste, ICTP Report
 No. IC/73/130, 1973 (unpublished).
 \bibitem{zavialov} O.I. Zavialov, {\sl Renormalized Quantum Field Theory}
(Kluwer
 Academic Press, 1990).
 \bibitem{inni} F. Barbero, A. Tiemblo and R. Tresguerres, Nuovo Cim. {\bf
 103A} (1990) 297; H. Dehnen, H. Frommert, Int. Jour. Th. Phys. {\bf 29} (1990)
 361; J.J. van der Bij, {\sl Bull. Board} hep-th\@xxx.lanl.gov - 9310064;
 see also A. Zee,
 Phys. Rev. Lett. {\bf 42} (1979) 417 and the review given in \cite{adler}.
 \bibitem{fr} M. Flato, R. R\c aczka, Phys. Lett. {\bf B208} (1988) 110.
 \end{thebibliography}
 \end{document}